\documentclass[secnumarabic,superscriptaddress,amssymb, nobibnotes, aps, prb]{revtex4-2}
\usepackage{geometry}
\usepackage{amsmath}
\usepackage{mathrsfs}  
\usepackage{mathtools} 
\usepackage{bbold}    
\usepackage{braket}    
\usepackage{gensymb}   

\usepackage{graphicx}      	

\usepackage{multirow}       
\usepackage{here} 			
\usepackage{xurl}		
\usepackage[plainpages=false,pdfpagelabels=true, linkcolor=cyan,breaklinks,colorlinks=true]{hyperref}  
\usepackage{soul}                   
\usepackage[dvipsnames]{xcolor}		
\usepackage{siunitx}
\usepackage{xspace}
\usepackage{booktabs}    

\newcommand{\ute}{UTe\textsubscript{2}\xspace}
\newcommand{\Tc}{$T$\textsubscript{c}\xspace}
\usepackage{graphicx}
\usepackage{float}  
\usepackage[utf8]{inputenc}

\begin{document}

\title{Giant transverse magnetic fluctuations at the edge of re-entrant superconductivity in \ute}%

\author{Valeska Zambra}
\affiliation{Institute of Science and Technology Austria, 3400 Klosterneuburg, Austria}
\author{Amit Nathwani}
\affiliation{Institute of Science and Technology Austria, 3400 Klosterneuburg, Austria}
\affiliation{California Institute of Technology, Pasadena, California 91125, USA}
\author{Muhammad Nauman}
\affiliation{Institute of Science and Technology Austria, 3400 Klosterneuburg, Austria}
\affiliation{Department of Physics and Astronomy, School of Natural Sciences (SNS), National University of Sciences and Technology (NUST), Islamabad 44000, Pakistan}
\author{Sylvia K. Lewin}
\affiliation{NIST Center for Neutron Research, National Institute of Standards and Technology, Gaithersburg, MD, USA}
\affiliation{Department of Physics, Quantum Materials Center, University of Maryland, College Park, MD, USA}
\author{Corey E. Frank}
\affiliation{NIST Center for Neutron Research, National Institute of Standards and Technology, Gaithersburg, MD, USA}
\affiliation{Department of Physics, Quantum Materials Center, University of Maryland, College Park, MD, USA}
\author{Nicholas P. Butch}
\affiliation{NIST Center for Neutron Research, National Institute of Standards and Technology, Gaithersburg, MD, USA}
\affiliation{Department of Physics, Quantum Materials Center, University of Maryland, College Park, MD, USA}
\author{Arkady Shekhter}
\affiliation{Los Alamos National Laboratory, Los Alamos New Mexico 87545, USA}
\author{B.~J.~Ramshaw}
\affiliation{Laboratory of Atomic and Solid State Physics, Cornell University, Ithaca, NY 14853, USA}
\affiliation{Canadian Institute for Advanced Research, Toronto, Ontario, Canada}
\author{K.~A.~Modic}
\email{kimberly.modic@ista.ac.at}
\affiliation{Institute of Science and Technology Austria, 3400 Klosterneuburg, Austria}

\begin{abstract}

\ute exhibits the remarkable phenomenon of re-entrant superconductivity, whereby the zero-resistance state reappears above 40 tesla after being suppressed with a field of around 10 tesla. One potential pairing mechanism, invoked in the related re-entrant superconductors UCoGe and URhGe, involves transverse fluctuations of a ferromagnetic order parameter. However, the requisite ferromagnetic order---present in both UCoGe and URhGe---is absent in \ute, and magnetization measurements show no sign of strong fluctuations. Here, we measure the magnetotropic susceptibility of \ute across two field-angle planes. This quantity is sensitive to the magnetic susceptibility in a direction transverse to the applied magnetic field---a quantity that is not accessed in conventional magnetization measurements. We observe a very large decrease in the magnetotropic susceptibility over a broad range of field orientations, indicating a large increase in the transverse magnetic susceptibility. The three superconducting phases of \ute, including the high-field re-entrant phase, surround this region of enhanced susceptibility in the field-angle phase diagram. The strongest transverse susceptibility is found near the critical end point of the high-field metamagnetic transition, suggesting that quantum critical fluctuations of a field-induced magnetic order parameter may be responsible for the large transverse susceptibility, and may provide a pairing mechanism for field-induced superconductivity in \ute.  

\end{abstract}

\maketitle

\section*{Introduction}

Understanding the connection between magnetism and superconductivity in \ute is key to determining its superconducting order parameters and pairing mechanisms. A compelling feature of \ute is the re-emergence of superconductivity at high magnetic fields in a ``halo" of field angles around the $b$-axis \cite{Lewin_2024}. This re-emergence is coincident with a metamagnetic transition to a spin-polarized state (see \autoref{fig:Magnetization}a). A remarkable feature of both the re-entrant and spin-polarized phases is that they occur in samples that are too disordered to exhibit zero-field superconductivity \cite{Wu_2024, Frank_2024}. This suggests that the pairing mechanism of the re-entrant superconducting phase may be related to the metamagnetic transition. 

A comparison can be made with the uranium-based superconductors UCoGe and URhGe, which also exhibit field re-entrant and field-reinforced superconductivity \cite{Levy_2005}. Unlike \ute, UCoGe and URhGe host ferromagnetic order that onsets at temperatures above the superconducting \Tc \cite{Aoki_2019b}. In these compounds, it has been suggested that a magnetic field applied perpendicular to the magnetic easy axis induces transverse fluctuations of the ordered moment that drive spin-triplet pairing \cite{Hattori_2013}. While \ute lacks ferromagnetic order at zero field, the fluctuations may be provided by a field-induced phase. However, such fluctuations have not been observed, calling for further study of the magnetism of \ute in high magnetic fields. 

Our study focuses on the field-angle phase diagram of \ute at 4 kelvin, where all superconducting phases are suppressed and where the magnetic response can be studied in isolation. We measure the magnetotropic susceptibility \cite{Modic_2018,Shekhter_2023}, which is sensitive to the magnetic susceptibility in a direction perpendicular to the applied magnetic field. We measure in pulsed magnetic fields up to 60 T, and in a full range of angles spanning both the $ac$- and $bc$-planes. We find evidence for transverse magnetic fluctuations in both planes, with particularly large transverse fluctuations in a region of the $bc$-plane that lies at the edge of all three superconducting phases.

\section*{Experiment}

\begin{figure}[htbp]
	\centering
	\includegraphics[width=0.8\linewidth, trim=0cm 0cm 0cm 0cm, clip=true]{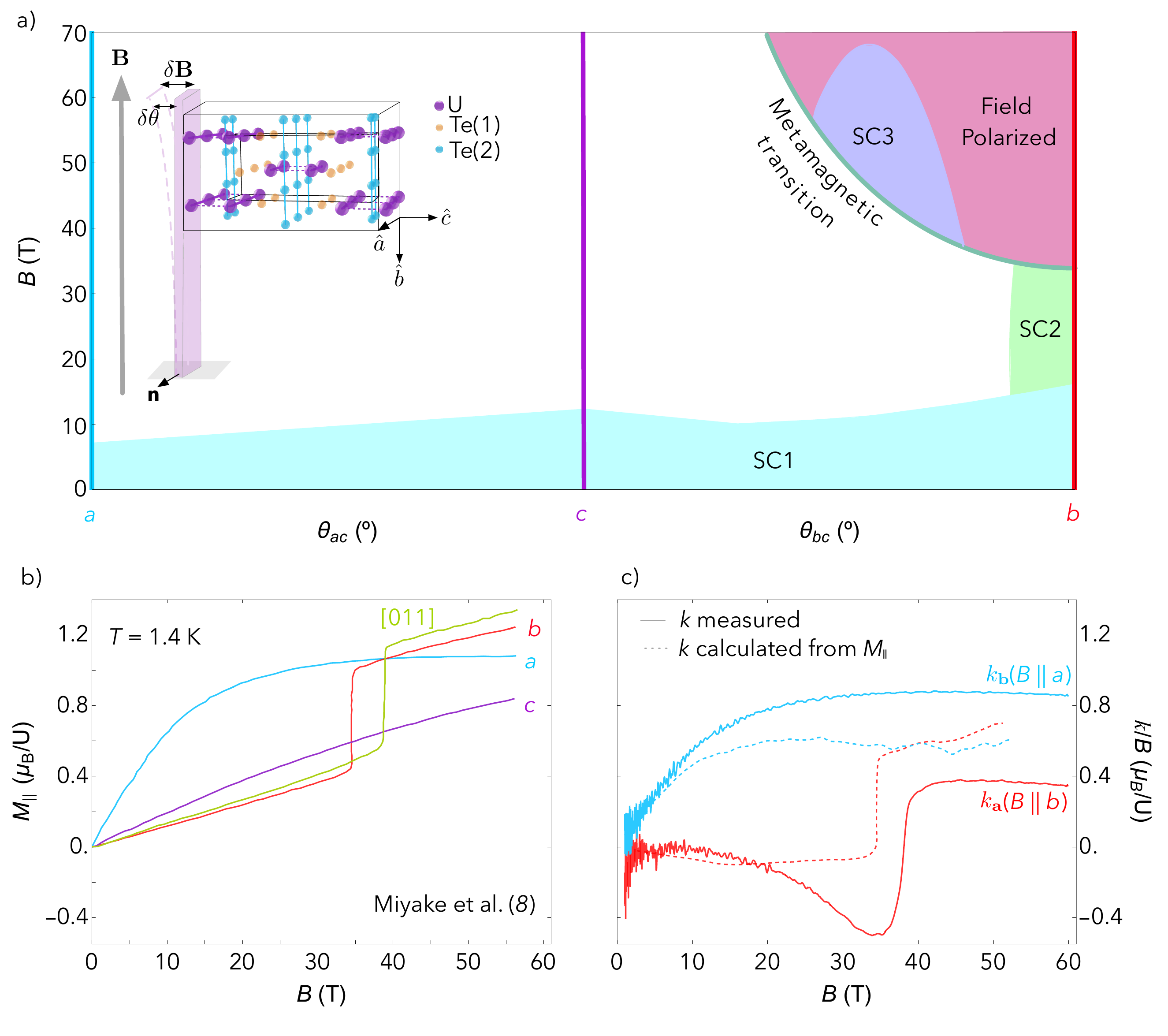}
	\rule{27em}{0.5pt}
	\caption[Crystal]{ \scriptsize \textbf{Phase diagram, magnetization, and magnetotropic susceptibility} a) The field-angle phase diagram of \ute at 300 mK, showing the field-angle phase boundaries of the SC1, SC2 and SC3 superconducting states (data reproduced from \citet{Helm_2024}, \citet{Knebel_2019}, and \citet{Ran_2019}). The colored vertical lines indicate directions where the data in panels b and c were measured. The inset shows the experimental geometry, where $\mathbf{B}$ is the applied magnetic field, and the transverse field component $\delta \mathbf{B}$ results from the vibration of the cantilever by the angle $\delta \theta$ around the axis $\mathbf{n}$. An example sample orientation is shown for measurements in the $bc$ plane, measuring $k_{\mathbf{a}}$ b) Magnetization as a function of magnetic field applied along each of the crystallographic directions, and in the [011] direction, at $T = 1.4$ K (data reproduced from \citet{Miyake_2021}). At low field, the $a$-axis is the magnetic easy axis. At the metamagnetic transition $B_\text{m} \approx 35$ T, the magnetization for field along the $b$-axis jumps to $\sim$1 $\mu_\text{B}$ per uranium and crosses the $a$-axis magnetization. c) The magnetotropic susceptibility divided by magnetic field $k /B$, measured at $T = $ 4 K, as a function of field for two different experiment geometries: with the oscillating field component $\delta \mathbf{B}$ in the $ac$-plane measuring $k_\textbf{b}$ (blue curve), and the oscillating field component $\delta \mathbf{B}$ in the $bc$-plane measuring $k_\textbf{a}$ (red curve). The subscripts on $k$ denote the normal vector $\textbf{n}$ to the vibration/rotation plane. The dashed lines are calculated  using \autoref{eq:magneto} and the measured magnetization from panel b (e.g. $k_\textbf{b}(\textbf{B}||b) = B\left(M_b-B\frac{\partial M_c}{\partial B_c}\right)$). The calculated magnetotropic susceptibility captures the overall qualitative behaviour of the measured magnetotropic susceptibility $k_\textbf{a}(B|| b)$ and $k_\textbf{b}(B||a)$---differences may be attributed to the presence of transverse susceptibility components that are not captured in the magnetization measurements.
	}
	\label{fig:Magnetization}
\end{figure}

We first describe the experimental geometry and why the magnetotropic susceptibility is sensitive to the transverse magnetic susceptibility. The inset of \autoref{fig:Magnetization}a shows the experimental geometry for a magnetotropic susceptibility measurement \cite{Modic_2018, Shekhter_2023}. A silicon microcantilever is driven near its fundamental bending mode, defining an axis $\textbf{n}$ around which the tip of the lever oscillates by a small angle $\delta \theta$. We place the cantilever in an external magnetic field $\textbf{B}$ and rotate the cantilever in the plane normal to the vector $\textbf{n}$. This constrains the field angle $\theta$ to always lie in the same plane as the lever oscillation angle $\delta \theta$.

The sample is placed on the tip of the cantilever with one crystallographic axis aligned along the length of the lever and another crystallographic axis aligned along $\textbf{n}$. For an orthorhombic crystal like \ute, this places the third axis perpendicular to the surface of the lever (inset of \autoref{fig:Magnetization}a). We perform two sets of rotation experiments: one in the $ac$-plane and one in the $bc$-plane. In all experiments reported here, the $c$-axis is perpendicular to the surface of the cantilever, and $\theta$ is defined as the angle between the applied magnetic field and the $c$-axis.

When a magnetic field is applied along one of the crystallographic axes of \ute, it produces a magnetic moment parallel to that axis. When the magnetic field is rotated away from that axis, the moment is no longer parallel to the magnetic field. This produces a magnetic torque, $\mathbf{\tau} = \mathbf{M} \times \mathbf{B}$, that bends the cantilever by a small angle. The angular derivative of this torque defines the magnetotropic susceptibility: $k \equiv \partial{\tau}/\partial \theta$. This susceptibility adds to the elastic bending stiffness of the cantilever and is measured by the shift in the cantilever resonance frequency (see SI for details of the calibration procedure). In terms of the vector $\mathbf{n}$ around which the cantilever oscillates in an applied field $\mathbf{B}$, the magnetotropic susceptibility is
\begin{align}
	k_\mathbf{n} (\mathbf{B}) = \mathbf{(n \times B)\cdot (n \times M)-\frac{1}{\mu_0}(n \times B) \cdot \chi (\mathbf{B}) \cdot (n \times B)},
	\label{eq:magneto}
\end{align}
where $\chi_{ij} (\mathbf{B}) \equiv \mu_0\partial M_i(\mathbf{B}) /\partial B_j$ is the differential magnetic susceptibility at applied field $\textbf{B}$ \cite{Shekhter_2023}. The first term in \autoref{eq:magneto} captures how the torque changes when a fixed moment $\textbf{M}$ is rotated in a field $\textbf{B}$ around the axis $\textbf{n}$. The second term captures how the torque changes due to the change in the moment itself as the crystal is rotated in the field. 

We can now illustrate why $k$ is sensitive to the \textit{transverse} magnetic susceptibility. The second term of \autoref{eq:magneto} selects the susceptibility tensor component that is perpendicular to both $\textbf{n}$ and $\textbf{B}$. For field along the $c$-axis, and with the sample oscillating in the $ac$-plane ($bc$-plane), this selects $\chi_{aa} (\textbf{B} || \textbf{c})$ ($\chi_{bb} (\textbf{B} || \textbf{c})$). These are what we define as transverse magnetic susceptibility components. Note that these are \textit{not} off-diagonal susceptibilities, such as $\chi_{bc} (\textbf{B} || \textbf{c})$, which are not allowed in an orthorhombic crystal structure for field along crystal axes. Instead, $\chi_{aa} (\textbf{B} || \textbf{c})$ and $\chi_{bb} (\textbf{B} || \textbf{c})$ are longitudinal (diagonal) susceptibility components that are measured perpendicular to the static, applied magnetic field. The oscillating field component perpendicular to the external field direction is generated by the oscillation of the cantilever (see $\delta \mathbf{B}$ in the inset of \autoref{fig:Magnetization}a). Further experimental details are given the Methods section.



\section*{Results}

\autoref{fig:Magnetization}c shows the measured magnetotropic susceptibility divided by magnetic field for two different crystal orientations on the cantilever, at 4 K, and with field applied along two different axes: $k_{\mathbf{a}}(\mathbf{B}||\mathbf{b})$ and  $k_{\mathbf{b}}(\mathbf{B}||\mathbf{a})$ (see SI for details of the sample orientations). \autoref{fig:Magnetization}b shows the measured magnetization along each of the principal crystallographic directions for comparison (reproduced from \citet{Miyake_2019}). The metamagnetic transition is clearly visible near 35 tesla for $\mathbf{B}||b$ in both the magnetization and the magnetotropic susceptibility measurements.

The magnetization in \autoref{fig:Magnetization}b is the \textit{longitudinal} magnetization: it is found by integrating the magnetic susceptibility measured \textit{along} the applied field direction, $M_i = \int \left(\partial M_i /\partial B_i\right) ~dB_i$ \cite{Miyake_2019}. We use the longitudinal magnetization and its field derivative---the longitudinal magnetic susceptibility---in conjunction with \autoref{eq:magneto} to calculate the expected magnetotropic susceptibility. This procedure does not account for any nonlinear transverse component to the magnetic susceptibility tensor (these components will be important later). This calculation is shown as dashed lines in \autoref{fig:Magnetization}c for $B||a$ and $B||b$. The overall magnitude and the qualitative features of the calculated and measured magnetotropic susceptibility are in good agreement. Small differences can be attributed to a small misalignment between the rotation vector $n$, the plane of the cantilever, and the sample's crystal axes. 
This demonstrates that, for field along the $a$- and $b$-axes, the measured magnetotropic susceptibility is largely determined by the longitudinal magnetic susceptibility. As shown below, this will not be the case for other field orientations.

\begin{figure}[htbp]
	\centering
	\includegraphics[width=0.95\linewidth, trim=0cm 0cm 0cm 0cm, clip=true]{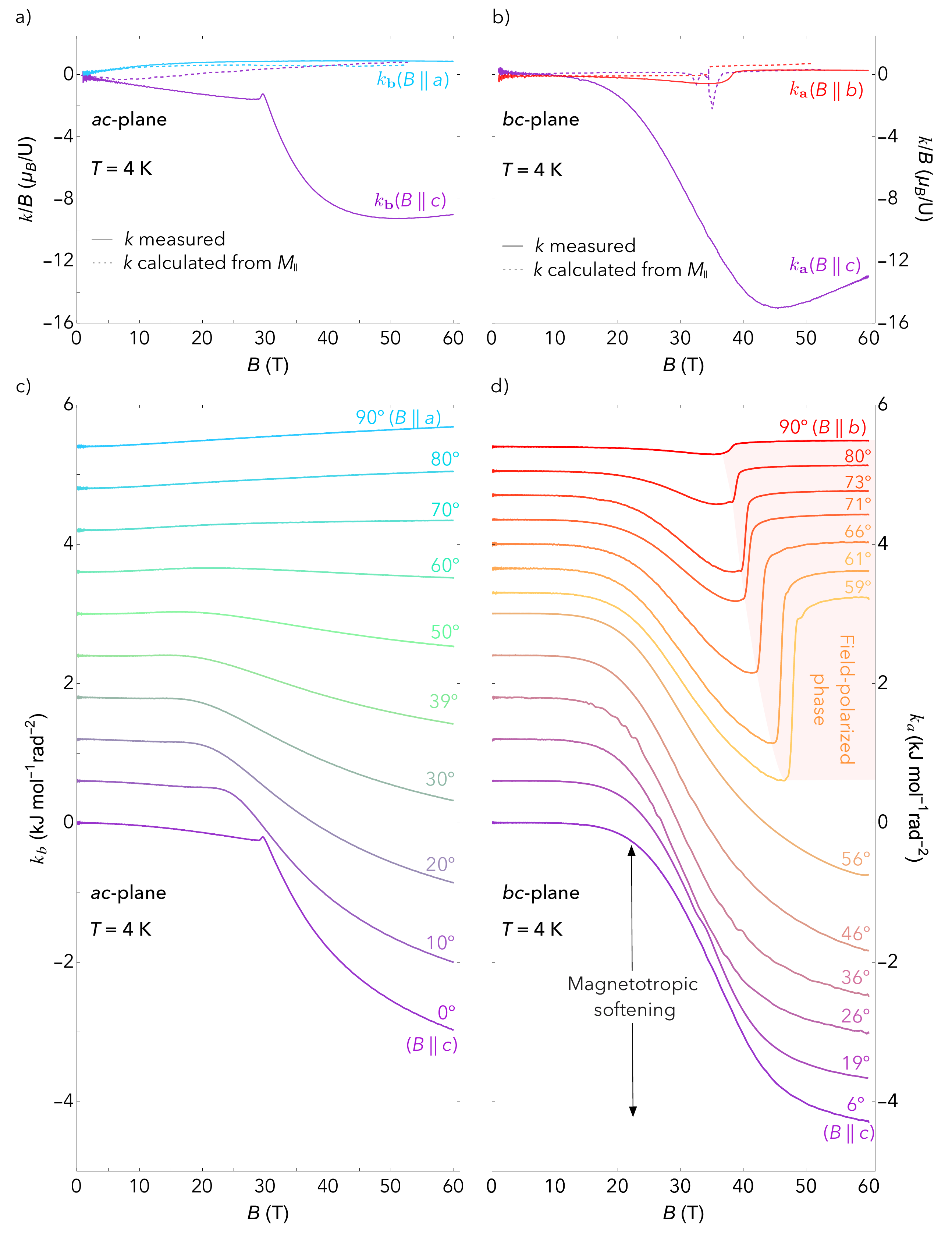}
	\rule{27em}{0.5pt}
	\caption[Crystal]{ \scriptsize \textbf{Angle-dependent magnetotropic susceptibility.} Magnetotropic susceptibility measurements performed in pulsed magnetic fields up to 60 T. Panel a) shows $k_b(B||c)$ and compares it to $k_b(B||a)$, and likewise for $k_a(B||c)$ and $k_a(B||b)$ in panel b). The dashed lines in both panels are the calculated values of $k$ based on the measured longitudinal magnetic susceptibility data from \autoref{fig:Magnetization}b and \autoref{eq:magneto} Panels c) and d) show the magnetotropic susceptibility measured at multiple angles in the $ac$- and $bc$-planes, respectively. As magnetic field approaches the $c$-axis in both planes, a large decrease is observed in $k$ that onsets at roughly 20 T. This decrease persists for a range of angles in both planes around $B||c$, and is abruptly cut off by the metamagnetic transition into the field-polarized phase (red shaded region in panel d) at $\theta = 59^\circ$ in the $bc$ plane.}
	\label{fig:Magnetotropic}
\end{figure}

\autoref{fig:Magnetotropic}a and b show the magnetotropic susceptibility for magnetic field applied along the $c$-axis, for both the $k_a$ and $k_b$ configurations. We also reproduce the data and calculations from \autoref{fig:Magnetization}c for comparison. Unlike the other two field orientations, the measured magnetotropic susceptibility for $\mathbf{B}||c$ deviates strongly from the estimate made using the longitudinal magnetization and susceptibility alone (dashed purple line). The large negative response in the magnetotropic susceptibility compared to that inferred from magnetization measurements indicates that a new susceptibility component, hidden from the longitudinal magnetization measurements, becomes active in the magnetotropic susceptibility at a field scale of around 20 tesla. 

Panels c and d in \autoref{fig:Magnetotropic} show the evolution of the magnetotropic susceptibility for a broad range of field angles in the $ac$- and $bc$-planes. The large decrease in the magnetotropic susceptibility that onsets near 20 T for field along the $c$-axis is observed over a broad range of angles in both planes. In the $bc$-plane (\autoref{fig:Magnetotropic}d), the decrease in the magnetotropic susceptibility is abruptly truncated by the metamagnetic transition into the field-polarized phase (red shaded region).

\section*{Analysis}

We uncover the origin of the large decrease in the magnetotropic susceptibility near 20 tesla by first analyzing the data along a high-symmetry direction. When the magnetic field is applied along a crystal axis, the first term in \autoref{eq:magneto} is completely determined by the longitudinal magnetization. Here, the second term has contributions only from the transverse magnetic susceptibility, i.e. $\chi_{ii}(B||j)$, for $j\neq i$. The longitudinal magnetization, shown in \autoref{fig:Magnetization}b, clearly shows no features that resemble the strong downturn seen in $k$ near 20 tesla. Therefore, the decrease in $k$ for field applied along the $c$-axis must originate from a transverse component of the magnetic susceptibility, i.e. $\chi_{ii}(B||j)$.

\begin{figure}[htbp]
	\centering
	\includegraphics[width=0.5\linewidth, trim=0cm 0cm 0cm 0cm, clip=true]{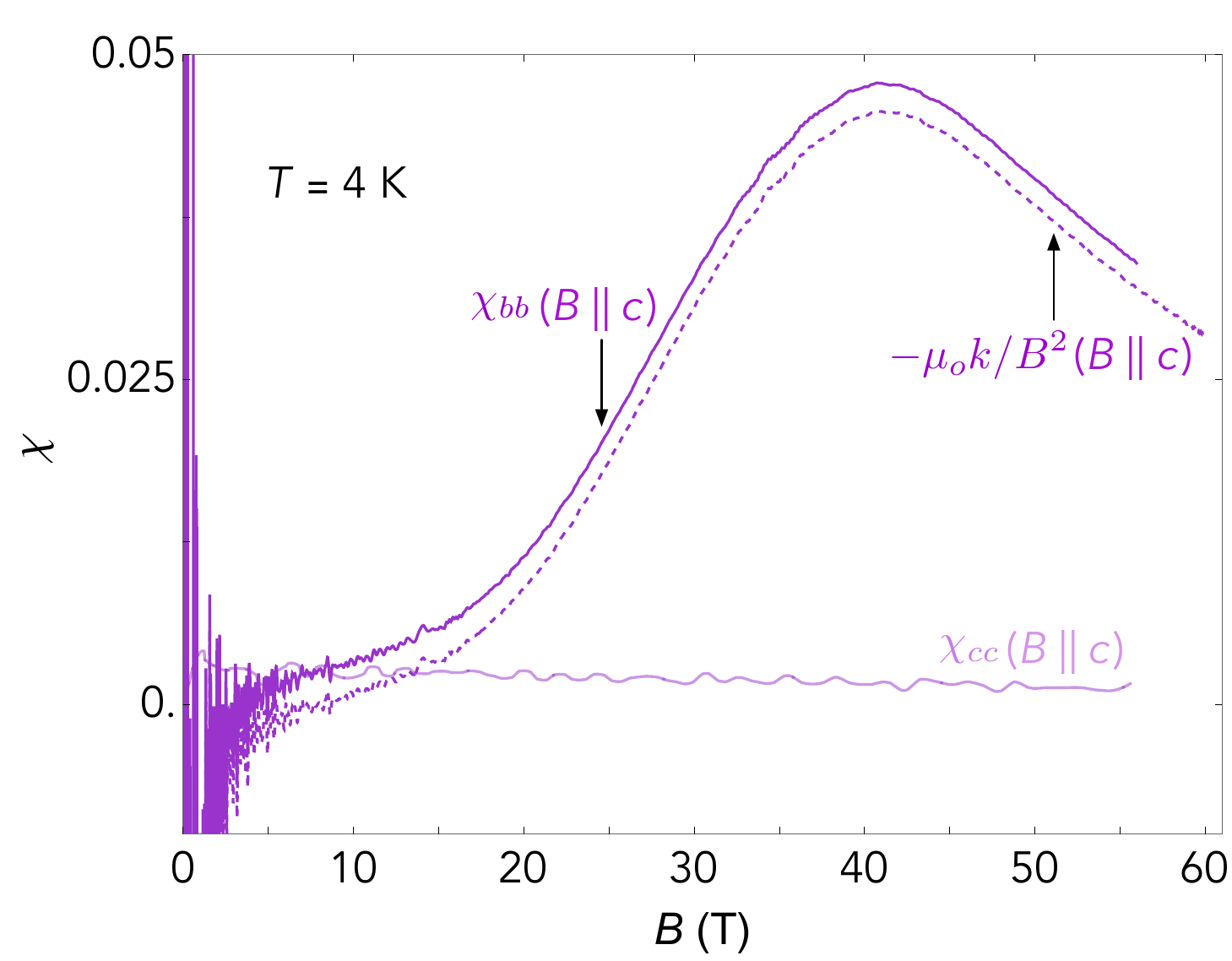}\\
	\rule{27em}{0.5pt}
	\caption[Crystal]{ \scriptsize \textbf{Longitudinal and transverse magnetic susceptibility} The dimensionless magnetic susceptibility as a function of magnetic field applied along the $c$-axis. The light purple curve shows $\chi_{cc} = \mu_o(\partial M_c / \partial B_c )$ obtained from conventional measurements of the longitudinal magnetic susceptibility \cite{Miyake_2021}. The dark purple curve shows the transverse magnetic susceptibility for $B||c$, taken as $\chi_{bb} = \mu_o (\chi_{cc}- k_a/B^2)$. } 
	\label{fig:Susceptibility}
\end{figure}

To highlight the large magnitude of the transverse magnetic susceptibility, we convert our magnetotropic data to dimensionless susceptibility units by dividing $k$ by $B^2/\mu_0$ (\autoref{fig:Susceptibility}). Next, we subtract out the contributions calculated using the longitudinal magnetization and magnetic susceptibility. The remaining contribution is the transverse magnetic susceptibility. \autoref{fig:Susceptibility} shows this susceptibility for $k_{a}$ measured with $B||c$, i.e. $\chi_{bb}(B||c)$. We also show the longitudinal susceptibility for the same field orientation, i.e. $\chi_{cc}(B||c)$. By 40 tesla, the transverse magnetic susceptibility is more than 30$\times$ larger than the longitudinal susceptibility; quantitatively, the transverse susceptibility is very large in comparison to the longitudinal susceptibility.

\begin{figure}[htbp]
	\centering
	\includegraphics[width=0.95\linewidth, trim=0cm 0cm 0cm 0cm, clip=true]{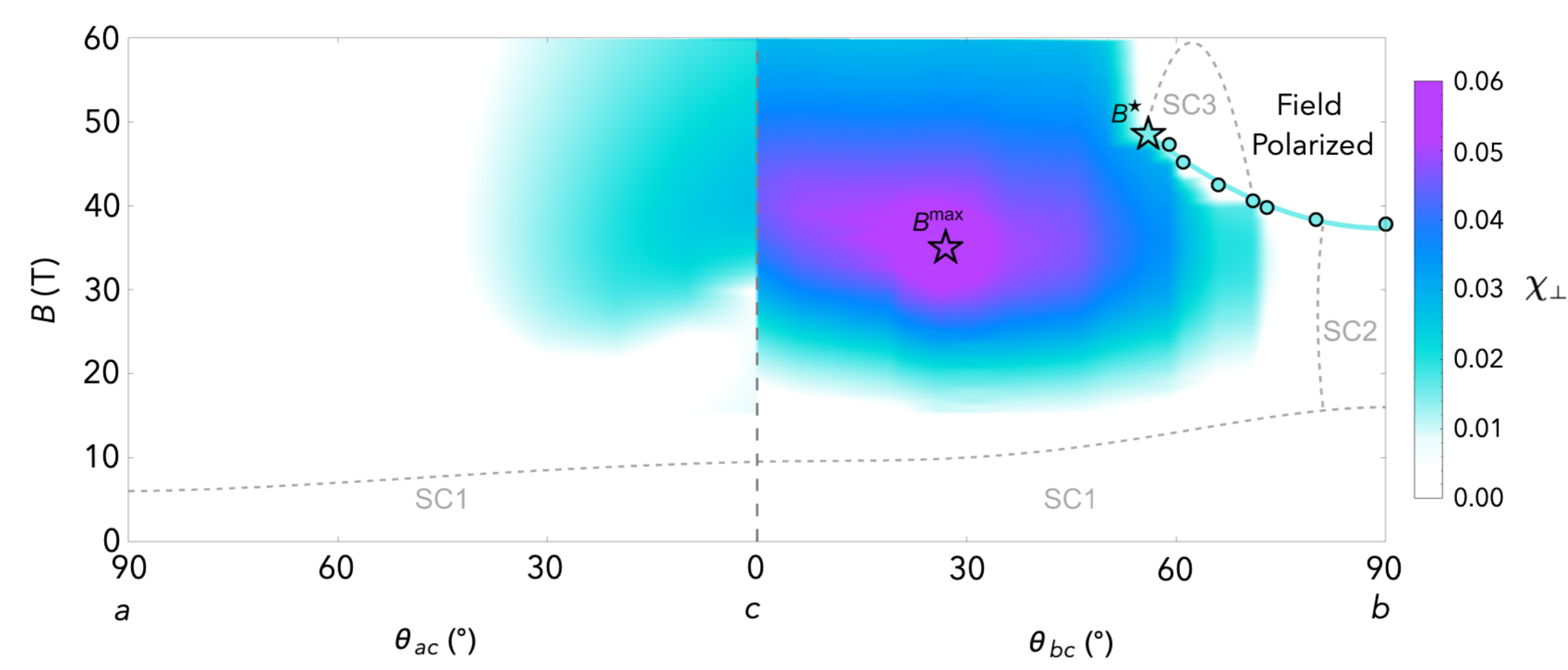}
	\rule{27em}{0.5pt}
	\caption[Crystal]{ \scriptsize \textbf{Phase diagram of the magnetotropic susceptibility, $k_\textbf{b}$ in the $ac$-plane and $k_\textbf{a}$ in the $bc$-plane, divided by $B^2$.} A color plot of the magnitude of the magnetotropic susceptibility converted to dimensionless susceptibility units as a function of field strength and field angle in the $ac$- and $bc$-plane at 4 kelvin. The measured magnetotropic susceptibility $k$ was multiplied by $-\mu_0/B^2$, which we define as $\chi_\perp$. As the longitudinal susceptibility components are not known for all intermediate angles, we cannot subtract those contributions from the measured signal. Rather, we assume, based on \autoref{fig:Susceptibility}, that the leading contribution is the transverse magnetic susceptibility. Magenta highlights the angle and field regions where the transverse magnetic susceptibility is largest. The values measured below 15 T are set to zero (white) on this plot because dividing by $B^2$ makes the noise diverge at low fields. The metamagnetic phase boundary as determined by the jump in $k$ (\autoref{fig:Magnetotropic}d) is indicated by teal points outlined in black and terminates at the critical endpoint near $\theta = 59^{\circ}$ as indicated by the teal star. The line through the metamagnetic transition points is a guide to the eye. The phase boundaries for SC1, SC2, and SC3 are taken from \citet{Lewin_2024} and \citet{Knebel_2019}.} 
	\label{fig:Unicorn}
\end{figure}

Moving away from the $c$-axis, the large increase in transverse susceptibility persists, and even strengthens, as we move towards the $b$-axis (\autoref{fig:Unicorn}). Like the longitudinal magnetization measured along the principal axes, the longitudinal magnetization measured at these intermediate angles indicates no substantial changes in the longitudinal susceptibility \cite{Miyake_2021}. Therefore, the large decrease in $k$ at all angles must come from an increase in the transverse magnetic susceptibility. While we do not have longitudinal magnetization measurements at all angles, and thus cannot subtract out the longitudinal component at all angles, \autoref{fig:Susceptibility} demonstrates that the transverse component is overwhelmingly larger than the longitudinal component in our measurement and that the transverse susceptibility is essentially equal to $-\mu_0 k/B^2$. We plot this quantity in \autoref{fig:Unicorn} for angles in the $ac$ and $bc$ planes. The transverse susceptibility is large for a broad region of the $bc$-plane, and peaks at a field of around 40 tesla and for angles between $10^\circ$ and $50^\circ$ away from the $c$-axis. 

\section*{Discussion}

\autoref{fig:Unicorn} shows a broad region of large transverse magnetic susceptibility in the $ac$- and $bc$-planes of the field-angle phase diagram. As shown in \autoref{fig:Susceptibility}, the transverse susceptibility reaches a value more than 30 times larger than the longitudinal susceptibility measured at the same field. A large transverse susceptibility implies large $\textit{fluctuations}$ of the magnetization in the direction transverse to the applied magnetic field, and large magnetic fluctuations are conducive to superconducting pairing. 

What could give rise to these large fluctuations? The most natural source would be soft magnons associated with a second-order magnetic phase transition \cite{Hattori_2013}. More specifically, the fact that we observe \textit{transverse} softening indicates that the phase transition is associated with the magnetic moment wanting to re-orient perpendicular to the applied field direction. For example, for $\mathbf{B}||c$, the large transverse susceptibility observed for both crystal orientations indicates that the field-induced moment wants to rotate away from the $c$ axis and into the $ab$-plane at high fields. The peak in $\chi_{\perp}$ at $B^{\rm max}$ in \autoref{fig:Unicorn} indicates the angle and field where the magnetic moment has the maximum tendency to reorientation toward the $\delta \textbf{B}$ we are probing.

What is the relationship between the large transverse susceptibility and the metamagnetic transition? \autoref{fig:Magnetotropic}d shows that the jump in the magnetotropic susceptibility across the metamagnetic transition increases in magnitude as magnetic field is rotated toward $B^\star$ between 56$^\circ$ and 59$^\circ$ in the $bc$-plane (teal star in \autoref{fig:Unicorn}). Likewise, \citet{Wu_2025} have shown that the discontinuity in the magnetization across the metamagnetic transition goes to zero upon approaching $B^\star$. Both of these observations are consistent with $B^\star$ marking the second-order critical end point of a line of first-order phase transitions. The fluctuations grow large in the vicinity of the metamagnetic phase boundary, and are abruptly truncated after the moment re-orients in the field polarized phase.

Why are the maximum transverse fluctuations not found right at $B^\star$? The reason is that our experiment is only sensitive to fluctuations in a direction perpendicular to $\mathbf{B}$, and the ferromagnetic order parameter does not necessarily grow in this direction at $B^\star$. In other words, there will be divergent magnetic fluctuations for \textit{some} field orientation at $B^\star$, but not in the direction we probe in this experiment. As noted by \citet{Wu_2025}, the critical endpoint at $B^\star$ forms a continuous line in field-angle space. Using the phase diagram of \citet{Wu_2025}, we find that the region of maximum transverse susceptibility in our experiment, denoted $B^{\rm max}$ in \autoref{fig:Unicorn}, is only $\approx 6^{\circ}$ away from $B^\star$ in another rotation plane. We suggest that at this $B^\star$, the magnetic order parameter grows along the direction $\delta \mathbf{B}$ we probe in our experiment---perpendicular to the applied field at $B^{\rm max}$. In the SI, we show data from a second sample oriented in a slightly different plane where we may indeed directly cross $B^\star$ in our measurement. 

Whatever their origin, strong ferromagnetic fluctuations are expected to be conducive to spin-triplet superconducting pairing \cite{Hattori_2013}. This pairing mechanism was suggested to explain the high-field superconductivity in UCoGe and URhGe \cite{Aoki_2019b, Miyake_2019}, which are easy-axis ferromagnets. We find evidence for strong transverse fluctuations over a broad region of the field-angle phase diagram (\autoref{fig:Unicorn}), and we suggest that they may be the ``glue'' for superconducting pairing in \ute. 

Measurements of the magnetotropic susceptibility are a new window into the high-field phase diagram of \ute. They go beyond traditional measurements of the longitudinal magnetization and susceptibility, and are compatible with pulsed magnetic fields and rotation studies. We reiterate that here we have only measured the transverse fluctuations for one particular sample-cantilever-field-rotation plane configuration. Future experiments will explore both different field-angle planes and different orientations of the $\textbf{n}$ vector, allowing us to probe different transverse susceptibility components and to determine the exact relationship between ferromagnetic fluctuations and the superconducting phases of \ute.

\section*{Acknowledgments}

We appreciate technical support from Ali Bangura and Zolt\'an K\"ollő.

V.Z., A.N., M.N. and K.A.M. acknowledge support from the European Research Council Starting Grant 101078696-TROPIC. V.Z., A.N., M.N. and K.A.M. thank the ISTA Nanofabrication Facility for technical support. B.J.R. acknowledges funding from the Office of Basic Energy Sciences of the United States Department of Energy under award number DE-SC0020143 for data analysis and writing. The National High Magnetic Field Laboratory is supported by the National Science Foundation through NSF/DMR-2128556*, the State of Florida, and the U.S. Department of Energy. A.S. acknowledges support from the DOE/BES “Science of 100 T” grant. A.S. thanks Downtown Subscription in Santa Fe, NM for their patience in hosting him. Sample preparation and characterization was supported by the NSF through DMR-2105191.

\section*{Author contributions}
B.J.R. and K.A.M. conceived of the experiment; S.K.L., C.E.F. and N.P.B. prepared and characterized the samples; V.Z., A.N., M.N. and A.S. performed the experiments and analyzed the data. V.Z., B.J.R., and K.A.M. wrote the manuscript with input from all co-authors.

\section*{Competing Interests}
The authors declare no competing interests.

\bibliographystyle{plainnat}
\bibliography{Main_arxiv.bib}

\begin{thebibliography}{14}
\providecommand{\natexlab}[1]{#1}
\providecommand{\url}[1]{\texttt{#1}}
\expandafter\ifx\csname urlstyle\endcsname\relax
  \providecommand{\doi}[1]{doi: #1}\else
  \providecommand{\doi}{doi: \begingroup \urlstyle{rm}\Url}\fi

\bibitem[Aoki et~al.(2019)Aoki, Ishida, and Flouquet]{Aoki_2019b}
Dai Aoki, Kenji Ishida, and Jacques Flouquet.
\newblock {Review of U-based Ferromagnetic Superconductors: Comparison between
  UGe2, URhGe, and UCoGe}.
\newblock \emph{Journal of the Physical Society of Japan}, 88\penalty0
  (2):\penalty0 022001, 2019.
\newblock ISSN 0031-9015.
\newblock \doi{10.7566/jpsj.88.022001}.

\bibitem[Frank et~al.(2024)Frank, Lewin, Salas, Czajka, Hayes, Yoon, Metz,
  Paglione, Singleton, and Butch]{Frank_2024}
Corey~E. Frank, Sylvia~K. Lewin, Gicela~Saucedo Salas, Peter Czajka, Ian~M.
  Hayes, Hyeok Yoon, Tristin Metz, Johnpierre Paglione, John Singleton, and
  Nicholas~P. Butch.
\newblock {Orphan high field superconductivity in non-superconducting uranium
  ditelluride}.
\newblock \emph{Nature Communications}, 15\penalty0 (1):\penalty0 3378, 2024.
\newblock \doi{10.1038/s41467-024-47090-1}.

\bibitem[Hattori and Tsunetsugu(2013)]{Hattori_2013}
K.~Hattori and H.~Tsunetsugu.
\newblock {p-wave superconductivity near a transverse saturation field}.
\newblock \emph{Physical Review B}, 87\penalty0 (6):\penalty0 064501, 2013.
\newblock ISSN 1098-0121.
\newblock \doi{10.1103/physrevb.87.064501}.

\bibitem[Helm et~al.(2024)Helm, Kimata, Sudo, Miyata, Stirnat, Förster,
  Hornung, König, Sheikin, Pourret, Lapertot, Aoki, Knebel, Wosnitza, and
  Brison]{Helm_2024}
Toni Helm, Motoi Kimata, Kenta Sudo, Atsuhiko Miyata, Julia Stirnat, Tobias
  Förster, Jacob Hornung, Markus König, Ilya Sheikin, Alexandre Pourret,
  Gerard Lapertot, Dai Aoki, Georg Knebel, Joachim Wosnitza, and Jean-Pascal
  Brison.
\newblock {Field-induced compensation of magnetic exchange as the possible
  origin of reentrant superconductivity in UTe2}.
\newblock \emph{Nature Communications}, 15\penalty0 (1):\penalty0 37, 2024.
\newblock \doi{10.1038/s41467-023-44183-1}.

\bibitem[Knebel et~al.(2019)Knebel, Knafo, Pourret, Niu, Vališka, Braithwaite,
  Lapertot, Nardone, Zitouni, Mishra, Sheikin, Seyfarth, Brison, Aoki, and
  Flouquet]{Knebel_2019}
Georg Knebel, William Knafo, Alexandre Pourret, Qun Niu, Michal Vališka,
  Daniel Braithwaite, Gérard Lapertot, Marc Nardone, Abdelaziz Zitouni, Sanu
  Mishra, Ilya Sheikin, Gabriel Seyfarth, Jean~Pascal Brison, Dai Aoki, and
  Jacques Flouquet.
\newblock {Field-Reentrant Superconductivity Close to a Metamagnetic Transition
  in the Heavy-Fermion Superconductor UTe2}.
\newblock \emph{https://doi.org/10.7566/JPSJ.88.063707}, 88, 5 2019.
\newblock \doi{10.7566/jpsj.88.063707}.
\newblock URL \url{https://journals.jps.jp/doi/abs/10.7566/JPSJ.88.063707}.

\bibitem[Lewin et~al.(2024)Lewin, Czajka, Frank, Salas, Yoon, Eo, Paglione,
  Nevidomskyy, Singleton, and Butch]{Lewin_2024}
Sylvia~K Lewin, Peter Czajka, Corey~E Frank, Gicela~Saucedo Salas, Hyeok Yoon,
  Yun~Suk Eo, Johnpierre Paglione, Andriy~H Nevidomskyy, John Singleton, and
  Nicholas~P Butch.
\newblock {High-Field Superconducting Halo in UTe\$\_2\$}.
\newblock \emph{arXiv}, 2024.
\newblock \doi{10.48550/arxiv.2402.18564}.

\bibitem[Lévy et~al.(2005)Lévy, Sheikin, Grenier, and Huxley]{Levy_2005}
F.~L\'evy, I.~Sheikin, B.~Grenier, and A.~D. Huxley.
\newblock {Magnetic Field-Induced Superconductivity in the Ferromagnet URhGe}.
\newblock \emph{Science}, 309\penalty0 (5739):\penalty0 1343--1346, 2005.
\newblock ISSN 0036-8075.
\newblock \doi{10.1126/science.1115498}.

\bibitem[Miyake et~al.(2019)Miyake, Shimizu, Sato, Li, Nakamura, Homma, Honda,
  Flouquet, Tokunaga, and Aoki]{Miyake_2019}
Atsushi Miyake, Yusei Shimizu, Yoshiki~J Sato, Dexin Li, Ai~Nakamura, Yoshiya
  Homma, Fuminori Honda, Jacques Flouquet, Masashi Tokunaga, and Dai Aoki.
\newblock {Metamagnetic Transition in Heavy Fermion Superconductor UTe 2
  Letters}.
\newblock \emph{Journal of the Physical Society of Japan}, 88:\penalty0 63706,
  2019.
\newblock \doi{10.7566/jpsj.88.063706}.
\newblock URL \url{https://doi.org/10.7566/JPSJ.88.063706}.

\bibitem[Miyake et~al.(2021)Miyake, Shimizu, Sato, Li, Nakamura, Homma, Honda,
  Flouquet, Tokunaga, and Aoki]{Miyake_2021}
Atsushi Miyake, Yusei Shimizu, Yoshiki~J. Sato, Dexin Li, Ai~Nakamura, Yoshiya
  Homma, Fuminori Honda, Jacques Flouquet, Masashi Tokunaga, and Dai Aoki.
\newblock {Enhancement and Discontinuity of Effective Mass through the
  First-Order Metamagnetic Transition in UTe2}.
\newblock \emph{Journal of the Physical Society of Japan}, 90\penalty0
  (10):\penalty0 103702, 2021.
\newblock ISSN 0031-9015.
\newblock \doi{10.7566/jpsj.90.103702}.

\bibitem[Modic et~al.(2018)Modic, Bachmann, Ramshaw, Arnold, Shirer, Estry,
  Betts, Ghimire, Bauer, Schmidt, Baenitz, Svanidze, McDonald, Shekhter, and
  Moll]{Modic_2018}
K.~A. Modic, Maja~D. Bachmann, B.~J. Ramshaw, F.~Arnold, K.~R. Shirer, Amelia
  Estry, J.~B. Betts, Nirmal~J. Ghimire, E.~D. Bauer, Marcus Schmidt, Michael
  Baenitz, E.~Svanidze, Ross~D. McDonald, Arkady Shekhter, and Philip~J.W.
  Moll.
\newblock {Resonant torsion magnetometry in anisotropic quantum materials}.
\newblock \emph{Nature Communications 2018 9:1}, 9:\penalty0 1--8, 9 2018.
\newblock \doi{10.1038/s41467-018-06412-w}.
\newblock URL \url{https://www.nature.com/articles/s41467-018-06412-w}.

\bibitem[Ran et~al.(2019)Ran, Liu, Eo, Campbell, Neves, Fuhrman, Saha, Eckberg,
  Kim, Graf, Balakirev, Singleton, Paglione, and Butch]{Ran_2019}
Sheng Ran, I-Lin Liu, Yun~Suk Eo, Daniel~J. Campbell, Paul~M. Neves, Wesley~T.
  Fuhrman, Shanta~R. Saha, Christopher Eckberg, Hyunsoo Kim, David Graf, Fedor
  Balakirev, John Singleton, Johnpierre Paglione, and Nicholas~P. Butch.
\newblock {Extreme magnetic field-boosted superconductivity}.
\newblock \emph{Nature Physics 2019 15:12}, 15:\penalty0 1250--1254, 10 2019.
\newblock \doi{10.1038/s41567-019-0670-x}.
\newblock URL \url{https://www.nature.com/articles/s41567-019-0670-x}.

\bibitem[Shekhter et~al.(2023)Shekhter, McDonald, Ramshaw, and
  Modic]{Shekhter_2023}
A.~Shekhter, R.~D. McDonald, B.~J. Ramshaw, and K.~A. Modic.
\newblock {Magnetotropic susceptibility}.
\newblock \emph{Physical Review B}, 108\penalty0 (3):\penalty0 035111, 2023.
\newblock ISSN 2469-9950.
\newblock \doi{10.1103/physrevb.108.035111}.

\bibitem[Wu et~al.(2024{\natexlab{a}})Wu, Weinberger, Hickey, Chichinadze,
  Shaffer, Cabala, Chen, Long, Brumm, Xie, Lin, Skourski, Zengwei, Graf,
  Sechovsky, Lonzarich, Valiska, Grosche, and Eaton]{Wu_2024}
Z~Wu, T~I Weinberger, A~J Hickey, D~V Chichinadze, D~Shaffer, A~Cabala, H~Chen,
  M~Long, T~J Brumm, W~Xie, Y~Lin, Y~Skourski, Z~Zengwei, D~E Graf,
  V~Sechovsky, G~G Lonzarich, 1~M Valiska, F~M Grosche, and A~G Eaton.
\newblock {Quantum critical fluctuations generate intensely magnetic
  field-resilient superconductivity in UTe2}.
\newblock \emph{arXiv}, 2024{\natexlab{a}}.
\newblock \doi{10.48550/arxiv.2403.02535}.

\bibitem[Wu et~al.(2024{\natexlab{b}})Wu, Weinberger, Hickey, Chichinadze,
  Shaffer, Cabala, Chen, Long, Brumm, Xie, Lin, Skourski, Zhu, Graf, Sechovsky,
  Lonzarich, Valiska, Grosche, and Eaton]{Wu_2025}
Z~Wu, T~I Weinberger, A~J Hickey, D~V Chichinadze, D~Shaffer, A~Cabala, H~Chen,
  M~Long, T~J Brumm, W~Xie, Y~Lin, Y~Skourski, Z~Zhu, D~E Graf, V~Sechovsky,
  G~G Lonzarich, M~Valiska, F~M Grosche, and A~G Eaton.
\newblock {A quantum critical line bounds the high field metamagnetic
  transition surface in UTe}.
\newblock \emph{arXiv}, 2024{\natexlab{b}}.
\newblock \doi{10.48550/arxiv.2403.02535v3}.

\end{thebibliography}
\newpage

\clearpage
\pagestyle{empty}

\begin{figure}[H]
	\centering
	\includegraphics[page=1, width=\textwidth]{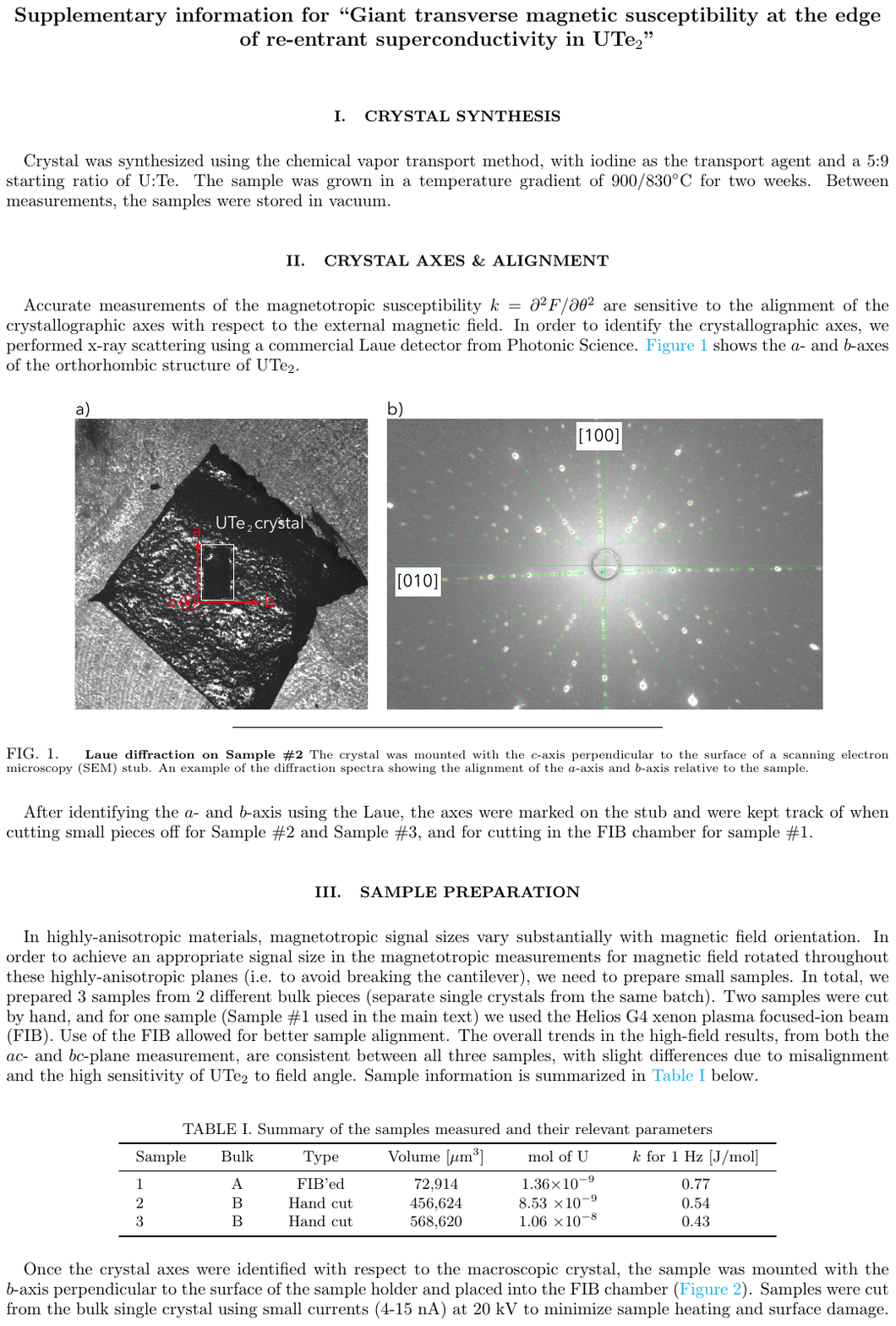}
\end{figure}

\begin{figure}[H]
	\centering
	\includegraphics[page=2, width=\textwidth]{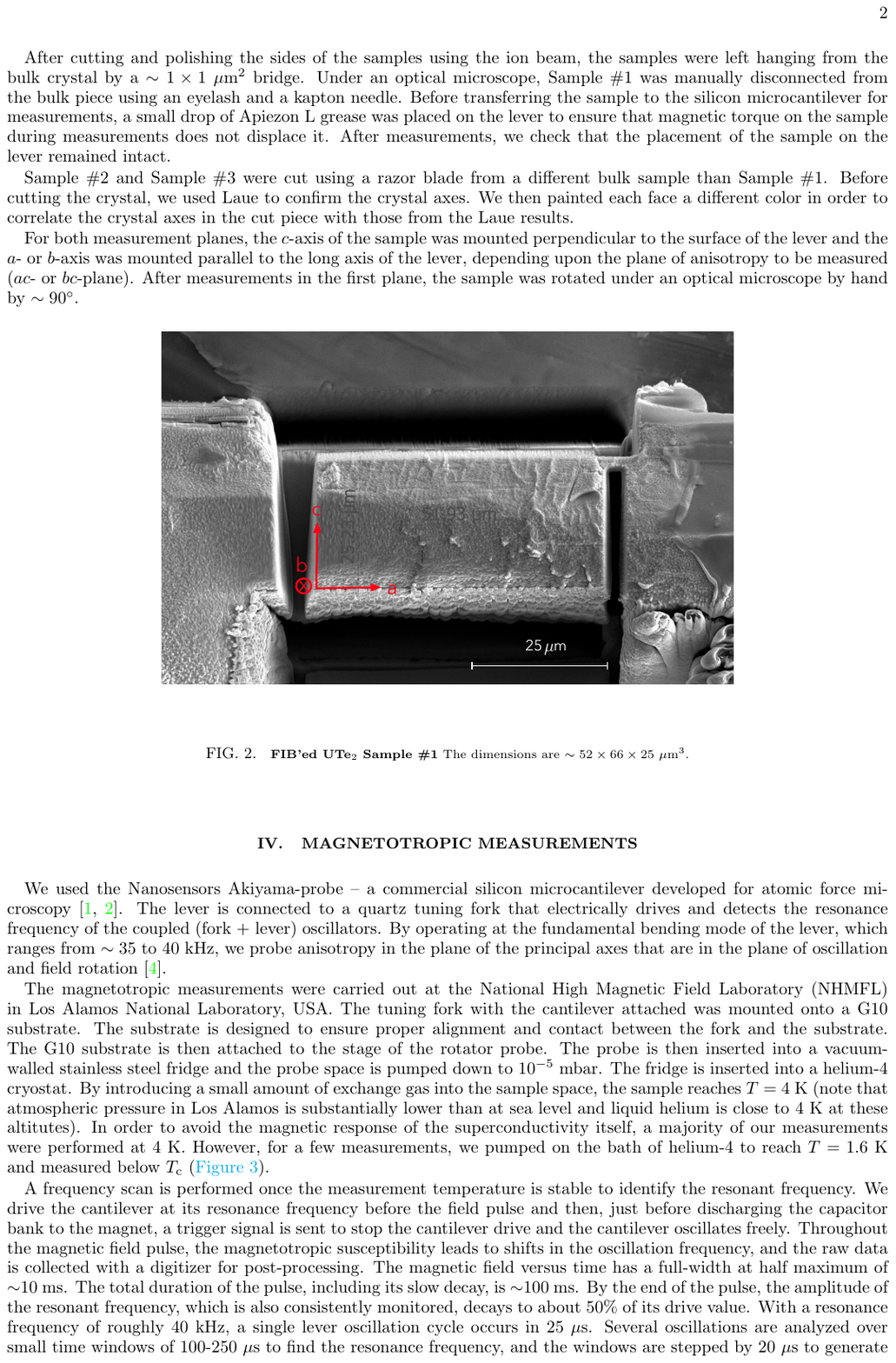}
\end{figure}

\begin{figure}[H]
	\centering
	\includegraphics[page=3, width=\textwidth]{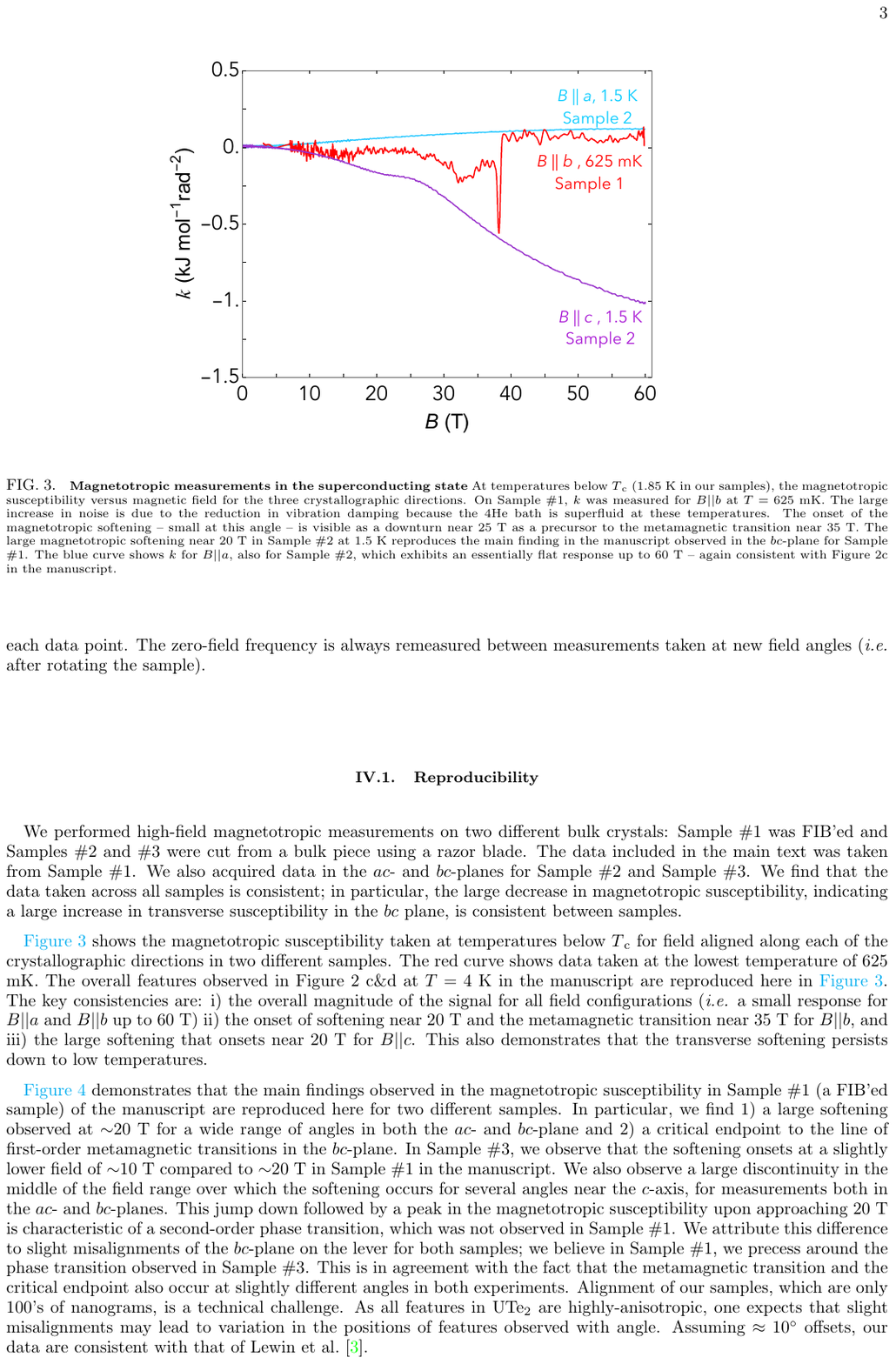}
\end{figure}

\begin{figure}[H]
	\centering
	\includegraphics[page=4, width=\textwidth]{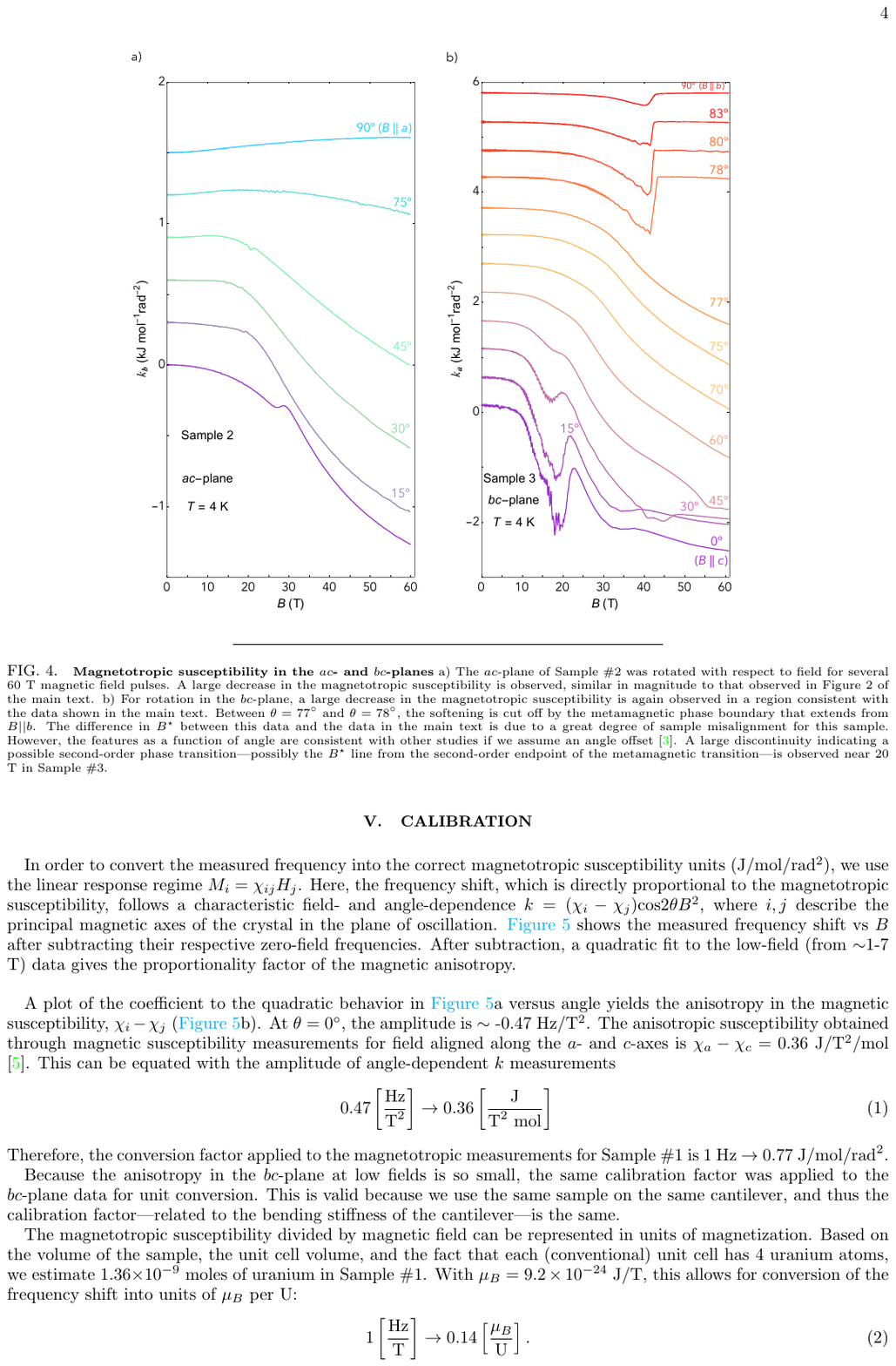}
\end{figure}

\begin{figure}[H]
	\centering
	\includegraphics[page=5, width=\textwidth]{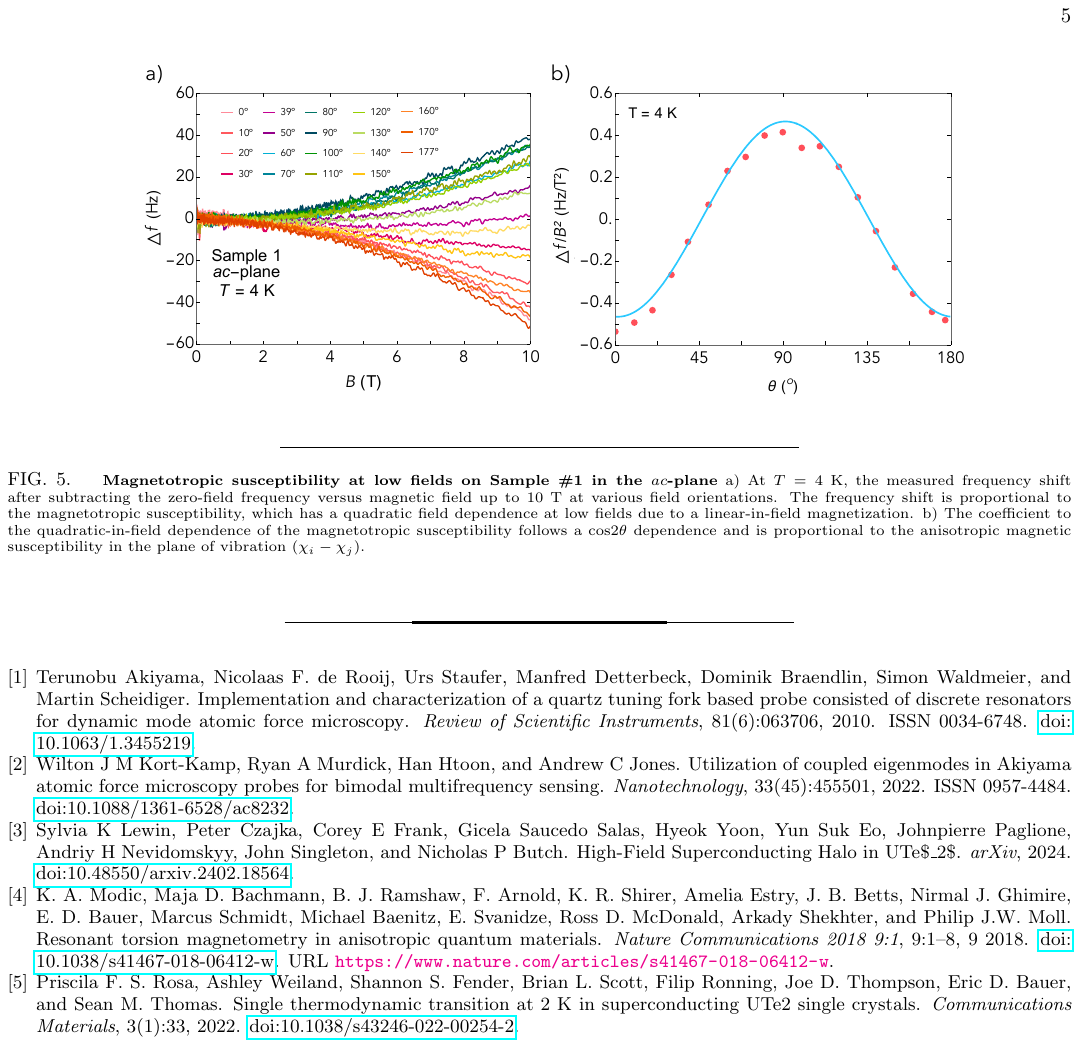}
\end{figure}

\end{document}